\documentclass{aa}

\usepackage{graphicx}
\usepackage{longtable}
\usepackage{colordvi}
\usepackage[varg]{txfonts}
\usepackage{txfonts}
\usepackage{hyperref}
\usepackage{color}
\usepackage{natbib}

\bibpunct{(}{)}{;}{a}{}{,} 

\begin{document}

   \title{Structural and dynamical modeling of WINGS
     clusters.}\subtitle{I. The distribution of cluster galaxies of
     different morphological classes within regular and irregular
     clusters}\author{A.~Cava \inst{1} \and A.~Biviano \inst{2} \and G.~A.~Mamon\inst{3} 
     \and J.~Varela\inst{4} \and D.~Bettoni\inst{5} \and M.~D'Onofrio\inst{6} 
     \and\\ G.~Fasano\inst{5} \and J.~Fritz\inst{7} \and M.~Moles\inst{4} 
     \and A.~Moretti\inst{5} \and B.~Poggianti\inst{5}}
 
   \offprints{A. Cava, \email{antonio.cava@unige.ch}} 
   \institute{Department of Astronomy, University of Geneva, 51 Ch. des Maillettes, 1290 Versoix, Switzerland
   \and INAF-Osservatorio Astronomico di Trieste ,Via Tiepolo 11, 34143 Trieste, Italy 
   \and Institut d'Astrophysique de Paris (UMR 7095: CNRS \& UPMC Sorbonne-Universit\'es), 98 bis Bd Arago, F-75014 Paris, France
   \and Centro de Estudios de F\'isica del Cosmos de Arag\'on, Plaza San Juan 1, E-44001 Teruel, Spain
   \and INAF - Osservatorio astronomico di Padova, Vicolo dell'Osservatorio 5, 35122 Padova, Italy
   \and Dipartimento di Fisica e Astronomia, Universit\`a degli Studi di Padova, Vicolo dell'Osservatorio 3, 35122 Padova, Italy
   \and Instituto de Radioastronom\'\i a y Astrof\'\i sica, UNAM, Campus Morelia, A.P. 3-72, C.P. 58089, Mexico
}
  \date{Received; accepted}
  
  \abstract {We study the distribution of galaxies in nearby
    clusters to shed light on the evolutionary processes at work
    within clusters and prepare for a full dynamical analysis to be
    conducted in forthcoming papers of this series.}{We use the
    Wide-field Nearby Galaxy-clusters Survey (WINGS) database
    complemented with literature data. We assign galaxy membership to
    individual clusters, then select a sample of 67 clusters with at
    least 30 spectroscopic members each. 53 of these clusters do not
    show evidence of substructures in phase-space, as measured by the
    Dressler-Shectman test, while 14 do. We estimate the virial radii
    and circular velocities of the 67 clusters by a variety of proxies
    (velocity dispersion, X-ray temperature, and richness) and use
    these estimates to build stack samples from these 53 and 14
    clusters, that we call `Reg' and `Irr' stacks, respectively. We
    show that our results are robust with regard to the choice of the virial
    radii and circular velocities used to scale galaxy radii and
    velocities in the stacking procedure. We determine the
    number-density and velocity-dispersion profiles of Elliptical (E),
    S0, and Spiral+Irregular (S) galaxies in the Reg and Irr samples,
    separately, and fit models to these profiles.}  {The number
    density profiles of E, S0, and S galaxies are adequately described by
    either a Navarro, Frenk, \& White (NFW) or a cored King model,
    both for the Reg and Irr samples, with a slight preference for the
    NFW model. The spatial distribution concentration increases from
    the S to the S0 and to the E populations, both in the Reg and the
    Irr stacks, reflecting the well-known morphology-radius relation.
    Reg clusters have a more concentrated spatial distribution of E
    and S0 galaxies than Irr clusters, while the spatial distributions of S galaxies in
    Reg and Irr clusters have a similar concentration.  We propose a
    new phenomenological model that provides acceptable fits to the
    velocity dispersion profile of all our galaxy samples.  The
    velocity dispersion profiles become steeper and with a higher
    normalization from E to S0 to S galaxies. The S0 velocity dispersion
    profile is close to that of E galaxies in Reg clusters, and intermediate
    between those of E and S galaxies in Irr clusters.}{Our results suggest
    that S galaxies are a recently accreted cluster population, that take $\leq
    3$ Gyr to evolve into S0 galaxies after accretion, and in doing so modify
    their phase-space distribution, approaching that of cluster
    ellipticals. While in Reg clusters this evolutionary process is mostly
    completed, it is still ongoing in Irr clusters.}{}
  \keywords{Galaxies: clusters: general -- Galaxies: clusters:
    kinematics and dynamics -- Galaxies: clusters: structure}
      
      \maketitle
      
      \section{Introduction}
\label{sec:intro}      
Clusters of galaxies have long been recognized as valuable tools for
the study of cosmology and galaxy formation. They are the most massive
virialized objects in the Universe and, as such, they sample the
rarest density fluctuations in the Universe, which makes them
excellent cosmological probes. They also provide excellent
laboratories for the study of the influence of environment on galaxy
evolution, on morphological evolution in particular. A useful tool in
this respect is the analysis of the spatial and velocity distribution
of cluster galaxies of different morphological types. In fact, a
galaxy morphological type can be affected by physical processes whose
efficiency depends both on the local density of galaxies or
intra-cluster diffuse hot gas, and on the galaxy peculiar velocity
within the cluster \citep[see, e.g.,][for a review of these
  processes]{Biviano11}.

In particular, the tidal damage (and stripping) consequent to
galaxy-galaxy collisions is stronger for low-velocity encounters,
eventually leading to a merger, and these encounters are more frequent
in high galaxy-density environments \citep{SB51}. Tidal effects and
mergers can transform the morphological type of a galaxy, most likely
destroying or thickening disks and creating bulges \citep{Barnes90}.  Repeated
high-speed encounters can have similar effects on not too massive
galaxies \citep{Moore+96}.  The interaction of a galaxy with the cluster
gravitational potential can lead to tidal truncation, depending on how
close the galaxy ventures to the cluster center \citep{Ghigna+98}.  Ram-pressure
stripping can remove the gas content of a galaxy, and this process is
more efficient in regions of high intra-cluster gas density, and for
galaxies moving at high speed in the cluster \citep{GG72}. Gas removal is
believed to favor the transformation from S to S0. 

There is strong observational evidence for morphological segregation
of cluster galaxies, both with respect to local density and
clustercentric distance, and with respect to velocity.  In nearby
clusters, Spiral and Irregular galaxies (S, hereafter) are known to
live in regions of lower local density \citep{Dressler80} and more
distant from the cluster center \citep{SS90b,WGJ93} than elliptical
galaxies (E, hereafter), with S0 galaxies (S0, hereafter) displaying
an intermediate spatial distribution between S and E
galaxies. Early-type Spirals live in regions of intermediate local
density between S0 and late-type Spirals \citep{TK06b}. Different
cluster galaxy populations also display different distributions in
velocity space, the velocity dispersion of E and S0 being smaller than
that of S \citep{Tammann72,MD77,Sodre+89}. Early-type Spirals have a velocity dispersion closer to S0 than to late-type Spirals
\citep{ABM98}.  Combining the spatial and velocity distribution in the
projected phase-space (PPS hereafter), \citet{Biviano+02} showed that
it is possible to distinguish three main cluster galaxy populations,
E+S0, early Spirals, and late Spirals.

At redshift $z \approx 0.5$, similar trends
of morphology with density and clustercentric radius were found in
clusters \citep{Postman+05,Smith+05}, but not in irregular ones
\citep{Dressler+97}. The morphological content of clusters changes
rapidly up to $z \approx 0.5$, with a decreasing fraction of S0 and an
increasing fraction of S with $z$
\citep{Dressler+97,Fasano+00,Postman+05,DOnofrio+15}, but then this evolution
slows down at higher $z$ \citep{Desai+07}. This change is stronger in
lower-mass clusters \citep{Poggianti+09}. The cluster velocity
dispersion is mildly dependent on the morphological fraction
\citep{Desai+07}, not only in low-$z$ but also in high-$z$ clusters,
up to $z \approx 1$ at least.

\citet{Fasano+15} have recently revisited the morphology-density
paradigm, using the WIde-field Nearby Galaxy-cluster Survey (WINGS)
database of galaxies in nearby clusters. They considered 5504
spectroscopically confirmed
members of 76 nearby clusters, down to an absolute magnitude limit
$M_V=-19.5$. \citet{Fasano+15} found
that the morphology-radius relation remains valid at all densities,
while the morphology-density relation is valid only in the inner
cluster regions, and outside the inner regions only for very regular
clusters without substructures.  This result does not depend on the
cluster mass, nor on the galaxy stellar mass. Galaxies of higher
stellar mass were found to display a stronger dependence of their
morphological type with density (but not with radius).

In this paper we extend the work of \citet{Fasano+15} by analyzing not
only the spatial, but also the velocity distributions of WINGS cluster
member galaxies of three morphological classes, E, S0, and S. Based on
the results of \citet{Fasano+15}, we use the clustercentric distance,
rather than the local density, to parametrize the spatial distribution
of different galaxy types. The morphological segregation in
clustercentric distance is analyzed by comparison of the number
density profiles (NDPs) of the three morphological classes.  In line
with previous studies \citep[e.g.,][]{ABM98,Desai+07}, the
morphological segregation in velocity is analyzed by comparison of the
velocity dispersion profiles (VDPs) of the three morphological
classes.  To take full advantage of the large statistical sample of
spectroscopically confirmed WINGS cluster members, we build two stacks
of the best-sampled WINGS clusters, one for the regular (REG) and
another for the irregular (IRR) clusters. The cluster regularity is
based on the presence of substructures detected with the method of
\citet[][DS method hereafter]{DS88}.  We will use the REG stack sample
and some of the results obtained in this paper in future papers
addressing the dynamics of the regular WINGS clusters (Mamon et
al. and Biviano et al., in preparation).

The structure of this paper is the following. We present the data
sample in Sect.~\ref{sec:data}. In Sect.~\ref{sec:stacks} we describe
our procedure to stack the data from different clusters.  The
PPS distributions (PPSD hereafter) are also presented
separately for E, S0, and S in the REG and IRR stacks. In
Sects.~\ref{sec:profs} and \ref{sec:vdps} we determine the NDPs and,
respectively, the VDPs, separately for E, S0, and S, in the REG and
IRR stacked clusters, to compare the relative spatial and velocity
distributions of these different morphological galaxy classes in
clusters of different dynamical status. Our results are discussed in
Sect.~\ref{sec:disc} and we provide our summary and conclusions in
Sect.~\ref{sec:concl}. Throughout this paper we assume a $\Lambda$
CDM cosmological model with $\Omega_\Lambda=0.7, \Omega_0=0.3,$ and
$H_0=70$ km~s$^{-1}$~Mpc$^{-1}$.

\section{The data}
\label{sec:data}
\subsection{The WINGS database}
\begin{figure}
\centering
\includegraphics[width=0.95\columnwidth]{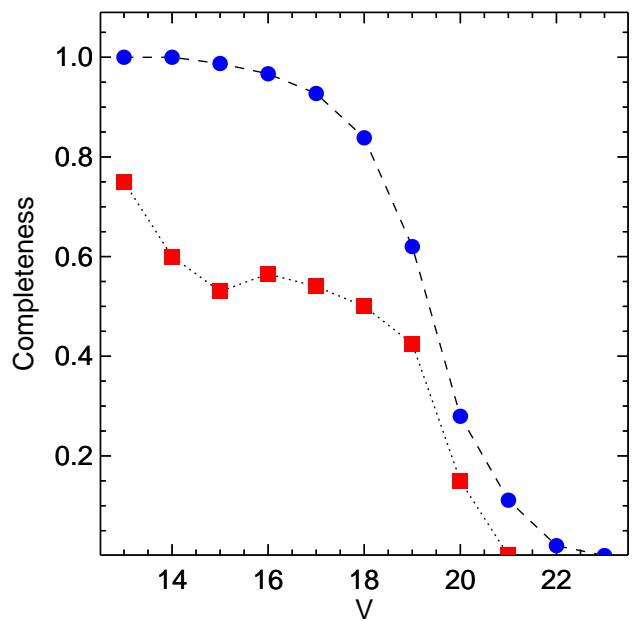} 
\caption{ Comparison of the V-band spectroscopic completeness for
  the original WINGS-SPE catalog presented in \citet{Cava+09} (red
  squares and dotted line; including 48 clusters, 6137 galaxies) and
  for the current WINGS+literature (blue dots and dashed line; 73
  clusters, 10277 galaxies) catalog.}
\label{fig:comp_compl}
\end{figure} 

WINGS \citep{Fasano+06} is an all-sky ($|b|>20$) survey of a complete,
X-ray selected sample of galaxy clusters in the redshift range
$0.04-0.07$. The goal of the WINGS project is the systematic study of
the local cosmic variance of the cluster population and of the
properties of cluster galaxies as a function of cluster properties and
local environment. The core of the WINGS project is the optical ({\it
  B}, {\it V}) imaging survey \cite[called WINGS-OPT;][]{Varela+09}.
It provides photometric data for large samples of galaxies
($\sim$550,000) and stars ($\sim$190,000) in the inner field
($34'\times34'$) of 76 nearby galaxy clusters, as well as structural and
morphological information for a subsample ($\sim$40,000) of
relatively bright galaxies \citep{Pignatelli+06,Fasano+12,DOnofrio+14}. Additional
photometric information comes from follow-up imaging surveys in the
near-infrared (NIR) {\it J} and {\it K} bands \cite[WINGS-NIR, 28 clusters;][]{Valentinuzzi+09}
and the {\it U} Johnson band \cite[17 clusters;][]{Omizzolo+14}. Currently, the
photometric data set is being expanded by very wide field observations
(four times the original WINGS area) taken with the VLT Survey Telescope (VST) OmegaCAM in the usual
{\it B} and {\it V} Johnson bands \citep{Gullieuszik+15} and in the
Sloan {\it u} band (D'Onofrio et al., in prep.).

The WINGS database is made available to the community through the
Virtual Observatory \citep{Moretti+14}.  WINGS is currently the only
data set providing a homogeneous database of detailed photometric,
morphological, and spectroscopic characteristics for several thousand
galaxies in galaxy clusters.

\subsection{Spectroscopic data: the extended (WINGS+literature) sample}
\label{ssec:spectro}
Spectroscopic information for $\sim6,000$ galaxies in the field of 48
WINGS clusters has been provided by a follow-up multi-fibre,
medium-resolution survey \cite[WINGS-SPE;][]{Cava+09}. Detailed
information about spectral features are available for a subset of
these galaxies \citep{Fritz+07,Fritz+11,Fritz+14}.  Follow-up
spectroscopy of a subset of the WINGS clusters has been obtained using the
Anglo Australian Telescope (AAT) AAOmega spectrograph \citep{Moretti+17}, 
to extend the spectroscopic coverage
of the WINGS clusters in their outer regions.  Detailed morphological
information is not yet available for the AAT/AAOmega additional
sample, so this is not considered in the present study.

We refer to \citet[][hereafter Paper~I]{Cava+09} for a detailed
description of the WINGS-SPE spectroscopic survey, including a
comparison with literature data. In this work we exploit the full
dataset, WINGS and literature data, to perform a detailed dynamical
analysis of the whole WINGS-OPT sample, including 76 galaxy clusters
(cluster A3562 is excluded because $V$-band images were obtained
under poor seeing conditions). In Paper~I we have already presented a
detailed data quality analysis and comparison with literature data,
detailing the method for including literature data in the WINGS-SPE
catalog \cite[see also][]{Moretti+14}. After updating the catalog with the
Sloan Digital Sky Survey (SDSS) DR7 and the latest  NASA/IPAC Extragalactic Database 
(NED) data, our final catalog includes $\sim10,000$
galaxies with redshift determinations ($\sim60\%$ from WINGS only),
increasing the global spectroscopic completeness level (defined as the number 
counts of galaxies with redshifts to the number
counts of galaxies in the photometric parent sample in magnitude bins, see Paper~I) 
for our sample of galaxy clusters from $\sim50\%$ to $\sim80\%$ for galaxies below 
magnitude $\sim 18$, and reaching $50\%$ completeness at a magnitude of $\sim 19.3$. 
The completeness improvement is remarkable, as shown in
Figure~\ref{fig:comp_compl}. In Paper~I, we compared multiple measurements
of galaxy velocities to estimate their average external error. We find
a rms $\sigma_{\Delta v} \sim90$ km~s$^{-1}$ in the worst case (see
Fig.~4 and Table~5 of Paper~I). This is larger than the average
internal error estimate ($\delta v\sim 45-50$~km~s$^{-1}$),
but still small enough not to affect the measurement of the internal
velocity dispersion of galaxy clusters.

Our final spectroscopic sample contains 10,288 galaxies in the field
of 76 galaxy clusters from the WINGS-OPT sample. The average number of
cluster members with redshifts is 93, rising to 105 for the clusters
in the WINGS-SPE sample, and lowering to 70 for the clusters outside
the WINGS-SPE sample. This difference reflects the higher level of
completeness attained by the WINGS multi-object spectroscopic
observations. In the following we concentrate on the subsample of 73
clusters with $\geq 10$ galaxies with measured spectroscopic redshifts
each, for a total of 10,277 galaxies.

We complement our sample with the photometric and morphological
information from the WINGS database \citep{Moretti+14}.  Galaxy
morphologies have been determined using the MORPHOT automatic tool of
\citet{Fasano+12}, which is optimized for the WINGS
survey. It applies maximum likelihood and neural network methods to
derive the best set of morphological parameters, and the equivalent
morphological index, $T_M$  (see Table~1 in \citealt{Fasano+12}
for a comparison between MORPHOT and visual classification
  indices).
Adopting the MORPHOT classification, we define the three subsamples
of ellipticals ('E'), with $T_M\leq -4$, lenticulars ('S0'), with
$-4<T_M\leq 0$, and spirals~+~irregulars ('S'), with $T_M>0$.

We finally assign a local completeness value to each galaxy in our
sample. We estimate the completeness of the spectroscopic sample by
comparing the number counts of galaxies with redshifts to the number
counts of galaxies in the photometric sample in magnitude bins. These
counts are estimated in radial bins around the cluster center, hence
our completeness is a function of both the galaxy magnitude and its
radial distance from the cluster center \citep[see][Equation~2]{Moretti+14}. 
In our completeness estimates
we also take into account that there are regions where there was no
observation at all due to geometrical constraints of the detectors,
for example, gaps or asymmetric positioning of the CCDs, and of the 
brightest cluster galaxy (BCG) with respect to the field center. 
Completeness corrections are required to correctly
estimate the intrinsic NDPs of the tracers of the
gravitational potential from the observed spatial distribution of
cluster members (see Sect.~\ref{sec:profs}).

\begin{figure}
\includegraphics[width=1.035\columnwidth]{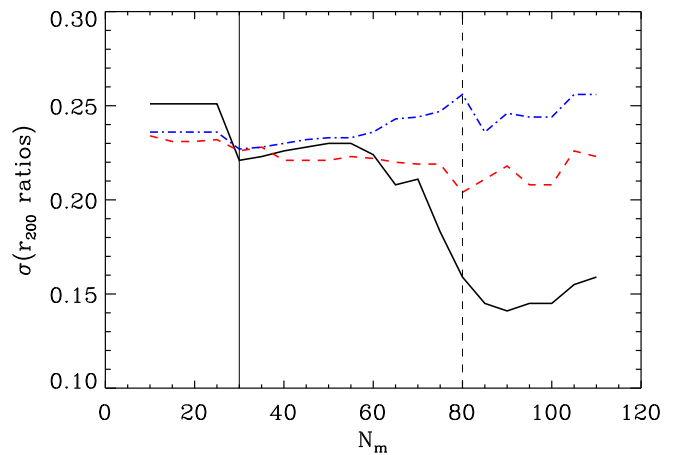} 
\caption{Dispersion of the values of
  $r_{200,\rm{X}}/r_{200,\sigma}$ (black solid line),
  $r_{200,N}/r_{200,\sigma}$ (red dashed line) and
  $r_{200,N}/r_{200,\rm{X}}$ (blue dash-dotted line) for samples of
  clusters with increasingly larger number of members, $N_{\rm m}$.  
}
\label{fig:r200xr200}
\end{figure} 

\section{The stacked samples of clusters}
\label{sec:stacks}
\subsection{Definition of the stacks}
The first step in our analysis is the cluster membership
determination. We adopt the ``Clean'' method of \cite{MBB13} to define cluster
members for each cluster in the spectroscopic parent sample based on
the projected phase-space ($R,v_z$) distribution. We call
$R$ (respectively $r$) the projected (respectively 3D) radial distance
from the cluster center, that we identify with the position of the BCG.

The rest-frame line-of-sight (LOS)
velocity is defined as $v_z=c \,
(z-\overline{z})/(1+\overline{z})$, where $z$ is the galaxy redshift,
$c$ is the speed of light, and $\overline{z}$ is the mean cluster
redshift, estimated at each new iteration of the membership
determination.

The Clean method works as follows:
\begin{enumerate}
\item look for gaps in the sorted velocity distribution using the gapper
  technique of \cite{WainerThissen76} with $C=4$ as proposed by
  \cite{Girardi+93};
\item estimate the virial radius and velocity ($r_{\rm v}$ and $v_{\rm v}$, respectively) from the aperture velocity
  dispersion ($\sigma_{\rm ap}$) computed with the robust median absolute deviation estimator,
  using a scaling of $\sigma_{\rm ap}/v_{\rm v}$ derived for Navarro, Frenk, and White (1996; NFW hereafter) models with
  velocity anisotropy estimated in \cite{MBM10};
\item \label{filter} filter the galaxies, keeping those within the estimated 
virial radius, $R<r_{\rm v}$, and within 2.7 predicted line-of-sight 
velocity dispersions for the NFW model with the \cite{MBM10} velocity
anisotropy from the median velocity,
  $|v_z-\overline{v}_z| < 2.7 \,\sigma_z^{\rm NFW}(R)$;
\item \label{rvirial} estimate the virial radius and velocity from the standard unbiased estimator of the
  aperture velocity dispersion within the previous estimated virial radius;
\item iterate on items~\ref{filter}--\ref{rvirial} until convergence on
  membership.
\end{enumerate} 

For a subset (50 clusters) of the WINGS parent cluster sample, also
X-ray temperatures, $T_X$, are available (extracted from the
BAX\footnote{BAX is the X-ray galaxy clusters database:
  {http://bax.ast.obs-mip.fr/.}} database). We provide alternative
estimates of the virial radii using the mass-temperature scaling relation
obtained from X-ray observations by
\citet{APP05}.  We call $r_{200,\rm{X}}$ these virial radius
estimates.

Yet another possibility is to use a richness-based estimate of the
virial radius (named `Num'; Mamon et al., in prep., see, \citealt{Old+14}).  
This is obtained by selecting cluster members in a rectangular area of 
projected phase space within a projected radius of 1 Mpc 
and an absolute line-of-sight velocity of $\pm1333$ km~s$^{-1}$ from
the cluster center in projected space and velocity.
 The richness is defined as the total number of members within this
 rectangular section of PPS, corrected for radial incompleteness. We
 calibrate the richness-based virial radius estimates using a robustly
 determined relation between the logarithm of the richness and the
 $\sigma_{\rm ap}$-based $\log r_{200,\sigma}$, and we refer to this
 richness-based estimate of the virial radius as $r_{200,N}$.

In the end, we have three estimates of the virial radius:
\begin{enumerate}
\item $r_{200,\sigma}$ or `sigv' from the cluster aperture velocity dispersion,
\item $r_{200,\rm{X}}$ or `temp' from the cluster X-ray temperature,
\item $r_{200,N}$ or `Num' from the richness.
\end{enumerate}
to which correspond three estimates of the circular velocity
$v_{200,\sigma}$, $v_{200,\rm{X}}$, and $v_{200,N}$, respectively.
For clusters without a X-ray temperature estimate, only the $r_{200,\sigma}$ and
$r_{200,N}$ are available. In Table~\ref{tab:clusters} we list
the main properties of the clusters in our sample, including the
available estimates of the virial radius.

\begin{table}
\centering
\caption{WINGS cluster properties.}
\label{tab:clusters}
\resizebox*{!}{0.95\textheight}{
\begin{tabular}{lrrrrrrr}
\hline 
Id & $\overline{z}$ & $N_{\rm{m}}$  & $f_{\rm{m},\rm{vir}}$ & $r_{200,\sigma}$ & $r_{200,\rm{X}}$ & $r_{200,N}$ & $P_{\rm{DS}}$ \\
\hline 
      A85 & 0.055 & 164 & 1.00 &  2.23 &  2.14 &  1.66  & 0.006 \\
     A119 & 0.044 & 175 & 1.00 &  1.87 &  2.01 &  1.78  & 0.030 \\
     A133 & 0.056 &  45 & 1.00 &  1.65 &  1.69 &  1.41  & 0.336 \\
     A147 & 0.044 &  21 & 1.00 &  1.42 &  --- &  1.31  & 0.913 \\
     A151 & 0.053 & 103 & 0.99 &  1.60 &  --- &  1.71  & 0.269 \\
     A160 & 0.044 &  81 & 1.00 &  1.52 &  0.99 &  1.60  & 0.324 \\
     A168 & 0.045 &  72 & 0.88 &  1.20 &  1.26 &  1.62  & 0.555 \\
     A193 & 0.048 &  73 & 1.00 &  1.60 &  1.24 &  1.59  & 0.840 \\
     A376 & 0.048 & 101 & 1.00 &  1.83 &  1.62 &  1.62  & 0.316 \\
     A500 & 0.068 &  92 & 0.99 &  1.57 &  --- &  1.70  & 0.358 \\
    A548b & 0.043 &  92 & 1.00 &  1.84 &  --- &  1.59  & 0.008 \\
     A602 & 0.060 &  51 & 1.00 &  1.49 &  --- &  1.40  & 0.345 \\
     A671 & 0.050 &  90 & 1.00 &  1.89 &  1.73 &  1.56  & 0.418 \\
     A754 & 0.055 & 230 & 1.00 &  2.23 &  2.48 &  1.84  & 0.138 \\
     A780 & 0.055 &  32 & 1.00 &  1.51 &  1.61 &  1.33  & 0.006 \\
    A957 & 0.045 &  88 & 1.00 &  1.52 &  1.31 &  1.52 & 0.096 \\
     A970 & 0.059 & 119 & 1.00 &  1.73 &  1.78 &  1.70  & 0.099 \\
    A1069 & 0.065 &  62 & 0.95 &  1.41 &  --- &  1.52  & 0.018 \\
    A1291 & 0.054 &  88 & 0.99 &  2.35 &  1.68 &  1.46  & 0.295 \\
   A1631a & 0.046 & 170 & 1.00 &  1.66 &  1.18 &  1.88  & 0.102 \\
    A1644 & 0.047 & 212 & 1.00 &  2.29 &  1.83 &  1.92  & 0.005 \\
    A1668 & 0.064 &  61 & 1.00 &  1.65 &  --- &  1.48  & 0.279 \\
    A1736 & 0.046 &  86 & 1.00 &  1.78 &  1.36 &  1.60  & 0.038 \\
    A1795 & 0.063 & 125 & 0.98 &  1.68 &  2.07 &  1.67  & 0.101 \\
    A1831 & 0.063 &  78 & 1.00 &  2.65 &  --- &  1.39   & 0.001 \\
    A1983 & 0.045 &  92 & 0.98 &  1.08 &  1.15 &  1.58  & 0.382 \\
    A1991 & 0.058 &  74 & 0.84 &  1.17 &  1.95 &  1.54  & 0.064 \\
    A2107 & 0.042 &  77 & 1.00 &  1.30 &  1.70 &  1.58  & 0.020 \\
    A2124 & 0.066 &  83 & 0.96 &  1.67 &  1.76 &  1.50  & 0.703 \\
    A2149 & 0.065 &  35 & 0.43 &  0.83 &  --- &  1.26  & 0.959 \\
    A2169 & 0.058 &  55 & 0.69 &  1.09 &  --- &  1.43  & 0.020 \\
    A2256 & 0.058 & 197 & 0.74 &  2.88 &  2.22 &  1.41  & 0.000 \\
    A2271 & 0.057 &  10 & 1.00 &  0.98 &  --- &  1.45  & 0.044 \\
    A2382 & 0.064 & 169 & 1.00 &  1.86 &  --- &  1.77  & 0.098 \\
    A2399 & 0.058 & 146 & 0.98 &  1.52 &  1.22 &  1.76  & 0.039 \\
    A2415 & 0.058 & 100 & 0.99 &  1.52 &  1.37 &  1.74  & 0.688 \\
    A2457 & 0.059 &  79 & 0.97 &  1.38 &  --- &  1.69  & 0.188 \\
   A2572a & 0.039 &  39 & 1.00 &  1.28 &  1.35 &  1.41 & 0.061 \\
    A2589 & 0.041 &  60 & 1.00 &  1.82 &  1.48 &  1.52  & 0.527 \\
    A2593 & 0.041 & 106 & 1.00 &  1.43 &  1.41 &  1.75  & 0.180 \\
    A2622 & 0.061 &  46 & 1.00 &  1.74 &  --- &  1.57  & 0.807 \\
    A2626 & 0.057 &  76 & 1.00 &  2.64 &  1.34 &  1.54  & 0.900 \\
    A2657 & 0.040 &  29 & 0.93 &  0.88 &  1.61 &  1.37  & 0.842 \\
    A2717 & 0.049 &  44 & 0.95 &  1.20 &  1.28 &  1.47  & 0.269 \\
    A2734 & 0.061 &  82 & 1.00 &  1.46 &  1.89 &  1.50  & 0.211 \\
    A3128 & 0.060 & 236 & 0.88 &  1.89 &  1.46 &  1.71  & 0.000 \\
    A3158 & 0.059 & 209 & 1.00 &  2.25 &  1.82 &  1.71  & 0.650 \\
    A3266 & 0.059 & 297 & 1.00 &  2.99 &  2.46 &  1.67  & 0.086 \\
    A3376 & 0.046 & 102 & 1.00 &  1.72 &  1.79 &  1.66  & 0.624 \\
    A3395 & 0.050 & 159 & 0.96 &  1.64 &  1.90 &  1.68  & 0.031 \\
    A3490 & 0.069 &  87 & 0.95 &  1.67 &  --- &  1.67  & 0.721 \\
    A3497 & 0.068 & 102 & 0.90 &  1.59 &  --- &  1.74  & 0.149 \\
   A3528a & 0.054 &  68 & 1.00 &  2.12 &  1.81 &  1.51  & 0.404 \\
   A3528b & 0.054 &  70 & 0.97 &  1.88 &  1.84 &  1.55  & 0.362 \\
    A3530 & 0.054 &  58 & 0.95 &  1.26 &  1.70 &  1.56  & 0.318 \\
    A3532 & 0.055 &  61 & 0.97 &  1.21 &  1.80 &  1.52  & 0.527 \\
    A3556 & 0.048 & 139 & 0.99 &  1.24 &  1.40 &  1.70  & 0.321 \\
    A3558 & 0.047 & 164 & 1.00 &  2.00 &  1.87 &  1.76  & 0.233 \\
    A3560 & 0.049 & 133 & 1.00 &  1.74 &  1.65 &  1.74  & 0.148 \\
    A3667 & 0.055 &  84 & 1.00 &  2.16 &  2.12 &  1.41  & 0.517 \\
    A3716 & 0.045 &  88 & 0.97 &  1.72 &  --- &  1.48  & 0.127 \\
    A3809 & 0.063 & 117 & 0.72 &  1.25 &  --- &  1.67  & 0.510 \\
    A3880 & 0.058 &  75 & 0.99 &  1.60 &  1.23 &  1.41  & 0.795 \\
    A4059 & 0.049 &  84 & 1.00 &  1.68 &  1.74 &  1.56  & 0.106 \\
  IIZW108 & 0.048 &  48 & 0.98 &  1.10 &  --- &  1.46  & 0.138 \\
    MKW3s & 0.045 &  61 & 1.00 &  1.28 &  1.67 &  1.55  & 0.978 \\
   RX0058 & 0.048 &  27 & 0.96 &  1.30 &  0.89 &  1.37 & 0.562 \\
   RX1022 & 0.054 &  55 & 1.00 &  1.51 &  1.07 &  1.29  & 0.201 \\
   RX1740 & 0.044 &  31 & 1.00 &  1.21 &  --- &  1.40  & 0.691 \\
    Z1261 & 0.065 &  13 & 0.85 &  1.14 &  --- &  1.14  & 0.949 \\
    Z2844 & 0.050 &  61 & 1.00 &  1.15 &  --- &  1.56  & 0.145 \\
    Z8338 & 0.049 &  74 & 0.95 &  1.54 &  --- &  1.59  & 0.574 \\
    Z8852 & 0.041 &  74 & 1.00 &  1.61 &  1.35 &  1.63  & 0.111 \\
\hline
\end{tabular}
}
\tablefoot{$N_{m}$ is the number of members and $f_{\rm m,vir}$ is the
  fraction of members at distances $\leq r_{200,\sigma}$ from
  the BCG centers.}
\end{table}

We build three stack samples of clusters by rescaling the projected
radial distances of cluster members by the cluster virial radii, $R_n
\equiv R/r_{200}$, and the rest-frame velocities by the cluster virial
velocities $v_n \equiv v_z/v_{200}$, one stack for each definition of
$r_{200}, v_{200}$ (from sigv, Num, and temp).

On the stack sample we run again the Clean procedure for interloper
removal. This is required because the Clean procedure can fail to reject
some interlopers when the size of the individual cluster sample is not
large enough. These remaining interlopers are, however, very few; less
than $\sim 1$\% of the initially selected members are identified as
interlopers by running the Clean procedure on the stack sample.

We do not consider all WINGS clusters in the stack, since some of them
do not have a sufficient number of members ($N_m$ hereafter) for a
reliable $r_{200}$ estimate. To choose the $N_m$ limit for inclusion
of a cluster in our sample, we have compared the values of
$r_{200,\sigma}$, $r_{200,\rm{X}}$ , and $r_{200,N}$. Decreasing
$N_m$ reduces the accuracy by which we can estimate $r_{200,\sigma}$ and
$r_{200,N}$.  In Fig.~\ref{fig:r200xr200} we show the dispersion
of the ratios $r_{200,\rm{X}}/r_{200,\sigma}$ (solid line),
$r_{200,N}/r_{200,\sigma}$ (dashed line), and
$r_{200,N}/r_{200,\rm{X}}$ (dash-dotted line) for samples of
clusters with increasingly larger $N_m$. Even if not strictly
monotonic, there is a decreasing trend for the dispersion of
$r_{200,\rm{X}}/r_{200,\sigma}$, that reaches a first (local) minimum
at $N_m \simeq 30$, and the absolute minimum at $N_m\simeq 90$. The
dispersion for the other $r_{200}$ ratios does not appear to depend on
$N_m$. We decide to consider the subsample of 68 clusters with $N_m
\geq 30$ and we check our results on a subsample of 38 clusters
with $N_m \geq 80$ (cutting at $N_m \geq 90$ would leave only 28
clusters, which we consider too small a sample to be representative of
the WINGS cluster population as a whole).

We run a 2D Kolmogorov-Smirnov (K-S) test \citep{FF87} to compare the
PPSDs  (the distributions in the $R_n, v_n$ space)
of the $N_m \geq 30$ and $N_m \geq 80$ stack samples, 
and we find no statistically significant difference for any scaling 
(sigv, Num, or temp; $p \geq 0.1$). Hence in the following we only
consider the $N_m \geq 30$ stacks, as they contain a larger data set
than the $N_m \geq 80$ ones.

\begin{figure}
  \centering
  \includegraphics[trim=2.cm 1.2cm 0.5cm 0.5cm, clip=true,width=1.\columnwidth]{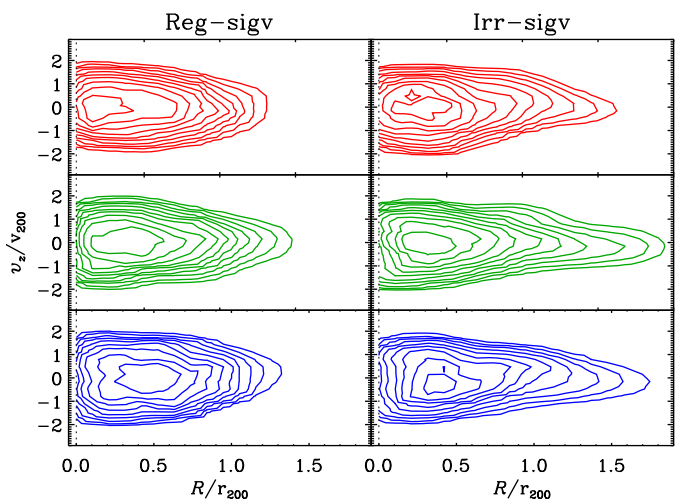}
  \caption{Adaptive-kernel density maps in projected phase-space
    (rest-frame velocities versus distances from the cluster centers) for
    the Reg-sigv (left panels) and Irr-sigv (right panels)
    stacks. Radii and velocities are in normalized units, using
    $r_{200,\sigma}$ and $v_{200,\sigma}$ for the normalization of
    radii and velocities, respectively.  Upper panel: E, middle panel:
    S0, lower panel: S.  Density contours are space logarithmically.}
\label{fig:rvstack}
\end{figure} 

\begin{figure}
\centering
  \includegraphics[trim=2.cm 1.2cm 0.5cm 0.5cm, clip=true,width=1.\columnwidth]{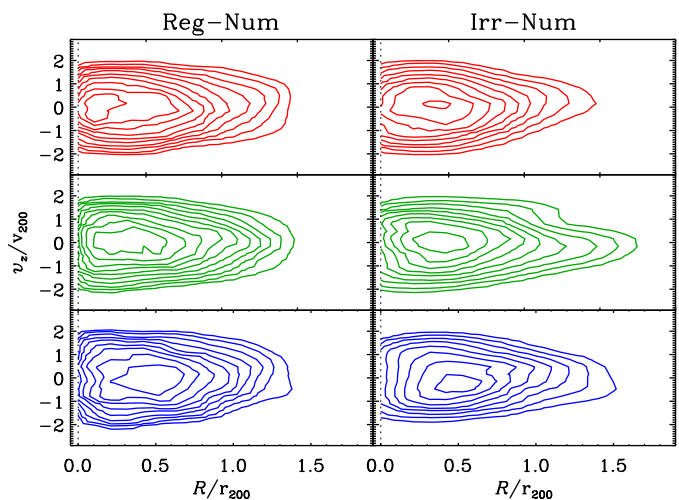}
\caption{Same as Fig.~\ref{fig:rvstack}, but for the Reg-Num (left
  panel) and Irr-Num (right panel) samples, and using $r_{200,N},
  v_{200,N}$ in lieu of $r_{200,\sigma}, v_{200,\sigma}$.}
\label{fig:rvstackN}
\end{figure} 

\subsection{Clusters with and without substructures} 
\label{ssec:subs}
We classify our sample into clusters with and without substructures,
based on the statistical test introduced by \citet[DS
  hereafter]{DS88}. The DS test evaluates the mean velocity
$\bar{\mathrm v}_{\rm local}$ and velocity dispersion $\sigma_{\rm
  v,local}$ of each group of $N_{\rm DS}$ neighboring cluster
members, and compares them with the corresponding quantities evaluated
for the whole cluster.  Rather than using the traditional DS method,
we adopt here the modified version of the DS method suggested by
\citet[][see also \citealt{Cava08_PhD}]{Biviano+02}. In short, this
modification consists in i) using $N_{\rm DS}=\sqrt{N_m}$ rather than
$N_{\rm DS}=11$, and ii) whenever a group has a velocity dispersion
larger than the cluster global velocity dispersion, this group is
not considered in the estimation of the global $\Delta$ parameter\footnote{The global $\Delta$ parameter is 
defined as the sum of the individual $\delta$'s of all galaxies, where $\delta$ is defined 
in Equation~1, Section~4.1 of \citet{Biviano+02}.} (since the
velocity dispersion is a proxy for the mass, this requirement means that we do not consider substructures that would be more massive than the
cluster in which they are embedded).

In Table~\ref{tab:clusters}, we list the probability values of the DS
statistic, $P_{\rm{DS}}$ (computed from 1000 resampling).  
We consider values $P_{\rm{DS}} \leq 0.05$
as significant evidence for substructure. Accordingly, we divide our
sample of 68 clusters with $N_m\geq30$ into two subsamples, the regular 
- `Reg' - sample of 54 clusters with $P_{\rm{DS}} > 0.05,$
and the irregular - `Irr' - sample of 14 clusters with
$P_{\rm{DS}} \leq 0.05$. From the two subsamples, we build six stacks,
two for each of our three scalings (sigv, Num, and temp), on which we
run the Clean algorithm to reject residual interlopers.

Using the 2D K-S test we compare the PPSDs (in normalized
units, $R_n, v_n$) of the Reg and Irr stacks obtained using the three
scalings and find no statistically different PPSDs
(probability $\geq 0.1$). This is also true of the Irr stacks obtained
using the temp and Num scalings. On the other hand, the Irr stack
obtained using the sigv scaling is significantly different from the
other two Irr stacks (probability $\leq 0.001$). For this reason, we
consider in the following both the stacks obtained using the sigv
scaling, that we name Reg-sigv and Irr-sigv, and the stacks obtained
using the Num scaling, that we name Reg-Num and Irr-Num (we prefer the
latter to the stacks obtained using the temp scaling because of better
statistics).

We show the PPSD of the Reg-sigv and Irr-sigv stacks in
Fig.~\ref{fig:rvstack}, and those of the Reg-Num and Irr-Num stacks in
Fig.~\ref{fig:rvstackN}, separately for E, S0, and S members. The 2D
K-S test indicates that E, S0, and S galaxies have significantly
(probability $<0.01$) different PPSDs in all four stacks,
even when restricting the comparison to the radial range covered by
both the populations being compared.

\section{The surface number density profiles} 
\label{sec:profs}
To obtain the number density profiles, $\Sigma(R)$, of our clusters we
use the photometric data rather than the spectroscopic data to avoid
the need for the correction for incompleteness of the spectroscopic
sample. We consider all galaxies in the cluster fields down to the
magnitude limit $V=19$, to ensure that we are using similar
populations of cluster galaxies for the determination of $\Sigma(R)$
and VDP, since most of our spectroscopic cluster members (that we use
for the determination of the VDP - see Sect.~\ref{sec:vdps}) are
brighter than $V=19$ (see Figure~\ref{fig:comp_compl}).  In
determining the $\Sigma(R)$ we correct the observed galaxy counts by
the fractions of the circular annuli covered by the imaging
observations.

We consider two models for the fit of $\Sigma(R)$,
\begin{enumerate}
\item the NFW model by
\citet{NFW96}, whose surface density profile was analytically derived by 
\citet{Bartelmann96}, as well as by \citet{LM01} in a slightly simpler form;
\item the \citet{King62} model, $\Sigma(R) \propto
  [1+(R/r_c)^2]^{-1}$.
\end{enumerate}
We add a constant surface density of background galaxies ($\Sigma_{\rm
  bg}$) to both models.  The NFW profile is theoretically motivated
and has been shown to fit the $\Sigma(R)$ of cluster galaxies
satisfactorily well \cite[see, e.g.,][]{BG03,LMS04}. The King model
differs from NFW in that it is characterized by a central core and a
sharper transition from the inner 3D density slope (zero for
King, $-1$ for NFW) to the outer slope of $-3$.

We use a maximum likelihood technique applied to the distribution of
projected radii to find the best-fit parameters of the two models,
namely the scale radius ($r_s$ in the case of the NFW model, and $r_c$
in the case of the King model), and the constant background surface
density, $\Sigma_{\rm bg}$, in units of the normalization of the
model. The normalization of the model is not a free parameter, as it
is constrained by the condition that the integral of the surface number 
density over the considered area is equal to the number
of observed galaxies (down to the chosen magnitude limit). We prefer
to leave $\Sigma_{\rm bg}$ a free parameter in the fit, rather than
using an average background density for all clusters, to account for
cosmic variance.

We limit the fit to the region between 50~kpc (basically excluding the
BCG) and ${\rm R_{\rm lim}}$, the radius of the most distant galaxy
from the BCG where the fraction of the area covered by the imaging
observations is $>0.6$. For most clusters ${\rm R_{\rm lim}} \geq 1$
Mpc, except for A168, where ${\rm R_{\rm lim}}<0.5$~Mpc. For this
reason, we exclude A168 from the following analysis, and consider only
the remaining 67 clusters from the $N_m \geq 30$ sample.

After fitting a model of $\Sigma(R)$ to the projected radii of all the
galaxies, we then fit separately the corresponding profiles of E, S0,
and S galaxies. In this second step of our procedure, we use a
single-parameter fit, and adopt the best-fit $\Sigma_{\rm bg}$ value
obtained by fitting $\Sigma(R)$ to the full galaxy set (hereafter, the
`All' sample), multiplied by the relative fractions of E, S0, and S
galaxies in the field. We take these fractions from Figs.~7 and 9 of
\citet{Bamford+09}, 0.1, 0.2, and 0.7 for E, S0, and S galaxies,
respectively. These values are based on the data set of $M_B \leq
-18.3$ galaxies of \citet[][after conversion to our adopted
  cosmology]{PG84}, Given typical $B-V$ galaxy colors
\citep[e.g.,][]{FSI95}, and the average redshift of galaxies in our
sample, $M_B=-18.3$ corresponds to an apparent magnitude limit $V \sim
18$, and a stellar mass limit $\log M_{\star} \sim 9.5$ \citep[using
  the relations of][]{BdJ01}.  The sample we use for the $\Sigma(R)$
determination is $\sim 1$ mag fainter that of \citet{PG84},
corresponding to a stellar mass $\log M_{\star} \sim 9.1$.  The field
morphological fractions change significantly with stellar mass, but
not below $\log M_{\star} \sim 9.5$, so the morphological fractions of
\citet{Bamford+09} are applicable to our sample.

\begin{figure}
     \includegraphics[trim=1.2cm 0.5cm 0.5cm 0.2cm, clip=true,width=1.0\columnwidth]{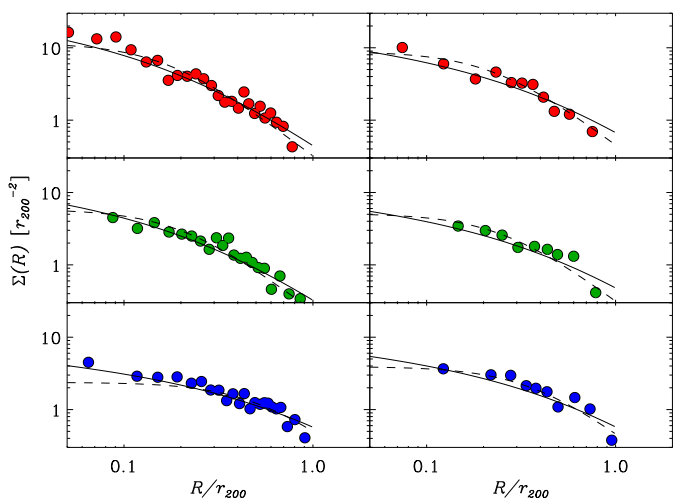}
   \caption{Projected surface number density profiles for E (Top
     panel), S0 (Middle panel), and S (Bottom panel) galaxies, in the
     Reg-sigv (left-hand panels) and Irr-sigv (right-hand
     panels) stacks, as determined using the spectroscopic samples of
     cluster members, and the completeness weights. Radii are in units
     of $r_{200,\sigma}$, $\Sigma(R)$ are in units of $r_{200,\sigma}^2$.
     The solid (respectively dashed) curves are the best-fit (projected) NFW
     (respectively King) profiles, as obtained by averaging the results of
     maximum likelihood fits to the $\Sigma(R)$ of individual clusters,
     obtained using the photometric sample.  Poisson error bars are
     smaller than the size of the symbols, but do not include the
     completeness uncertainties.}
\label{fig:densprof}
\end{figure} 
\begin{figure}
  \includegraphics[trim=1.2cm 0.5cm 0.5cm 0.2cm, clip=true,width=1.0\columnwidth]{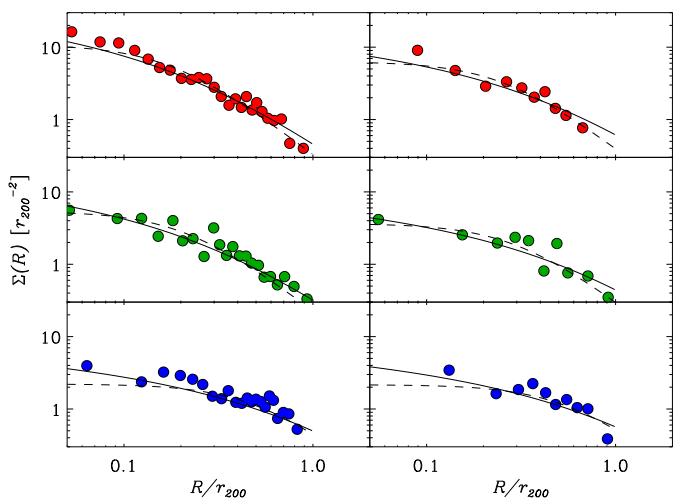}
   \caption{Same as Fig.~\ref{fig:densprof}, but for the Reg-Num and
     Irr-Num samples, using $r_{200,N}$ in lieu of
     $r_{200,\sigma}$. }
\label{fig:densprofN}
\end{figure} 

The statistics are good enough to allow fitting 47/42/33 of the 67
individual cluster E/S0/S surface number density profiles.  Using the
Bayesian information criterion
\begin{equation}
{\rm BIC} = -2\,\ln {\cal L}_{\rm max} + (\ln N_d)\,N_p \ ,
\end{equation}
where $N_d, N_p$ are the number of data and free parameters, respectively,
we establish that the projected NFW model provides a
better fit than the King model to 84, 78, and 67\% (respectively 80, 63, and
56\%) of the E, S0, and S galaxy surface density profiles for the Reg
(respectively Irr) cluster set. In general, the spatial distribution of
galaxies is therefore better described by the NFW model than by the
King model, but this is less commonly the case for galaxies of later
types and in irregular clusters.

To define the model profile to be used for the stacks, we then average
the values of the scale radius parameter of each cluster, in units of
the cluster $r_{200}$ ($r_{200,\sigma}$ or $r_{200,N}$), using
the number of cluster members as weights. While these values are
obtained using a subset of all the clusters that contribute to the
stack (i.e., those with sufficient statistics for allowing the $\Sigma(R)$
determination and fit), we can assume that each average is
representative of the full sample. We expect this assumption not to
introduce any significant bias in our analysis, given that clusters
with poor statistics (few member galaxies) do not contribute much to
the stacks anyway.

In Figs.~\ref{fig:densprof} and \ref{fig:densprofN} we compare the
average best-fit (projected) NFW and King models to the binned
$\Sigma(R)$, obtained from the four spectroscopic stack samples
Reg-sigv, Irr-sigv, Reg-Num, and Irr-Num.  In estimating the
stack binned $\Sigma(R)$ we take into account two completeness
corrections, namely we correct for the incompleteness of the spectroscopic sample, and for the fact that, in the outer regions, at a given radius only a
subset of clusters contribute to the stack. This second correction is
described in \citet{MK89,Biviano+02}. 
The normalization of the $\Sigma(R)$ model fits are
adjusted to fit the binned $\Sigma(R)$.  Overall, the visual
impression is that of a good agreement between the binned and fitted
$\Sigma(R)$, considering that they have been obtained by very
different procedures and that the completeness corrections for the
spectroscopic samples are very uncertain at large radii. 

\begin{figure}
  \includegraphics[trim=1.2cm 0.5cm 0.7cm 0.5cm, clip=true,width=1.0\columnwidth]{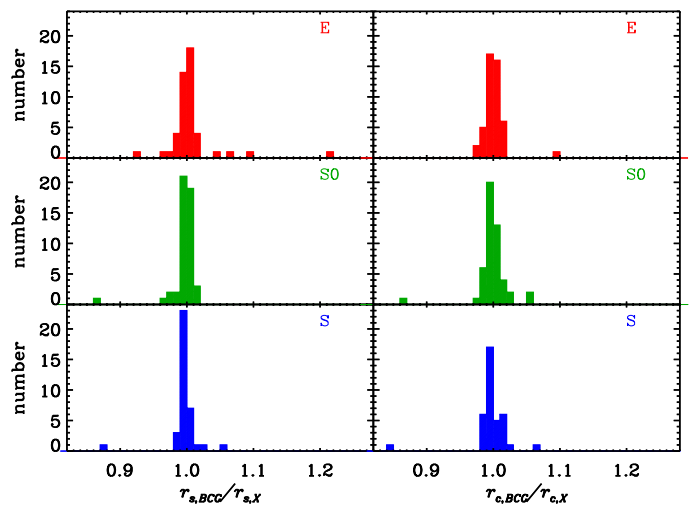}
  \caption{Histogram distributions of the ratio of best-fit NFW
       (left panels) and King (right panels) model scale
       radius parameters for individual clusters, using BCG and X-ray
       centers, for the E, S0, and S populations in the top, middle, and
       bottom panels, respectively.}
\label{fig:compCent}
\end{figure}

To estimate the effect of adopting the BCG position as a cluster
  center, we repeated our analysis by adopting an alternative center
  definition, namely the peak of the cluster X-ray emission.  The
  average distance of the X-ray and BCG centers for our clusters is
  $\sim 100$~kpc. This difference does not result in a significant
  difference of the PPSDs of the stacked samples, according to a 2D
  Kolmogorov-Smirnov test, for any of the morphological classes and
  adopted scalings. Changing the center definition has also a
  negligible effect on the membership determination for the different
  stacks. A more quantitative comparison is provided by
  Fig.~\ref{fig:compCent}, where we show the distribution of the
  ratios of the best-fit scale radii of our clusters, obtained using
  the two center definitions, for the three morphological classes and
  both the NFW and King models. Most of the ratios are within 0.03 of
  unity. As a result, the effect of a different center choice on the
  stacked NDPs is even smaller. We also checked that the different
  center choice has little effect on the VDPs (described in
  Sect.~\ref{sec:vdps}).

\begin{figure}
   \includegraphics[trim=1.cm 0.75cm 1.25cm 0.5cm,
     clip=true,width=1.0\columnwidth]{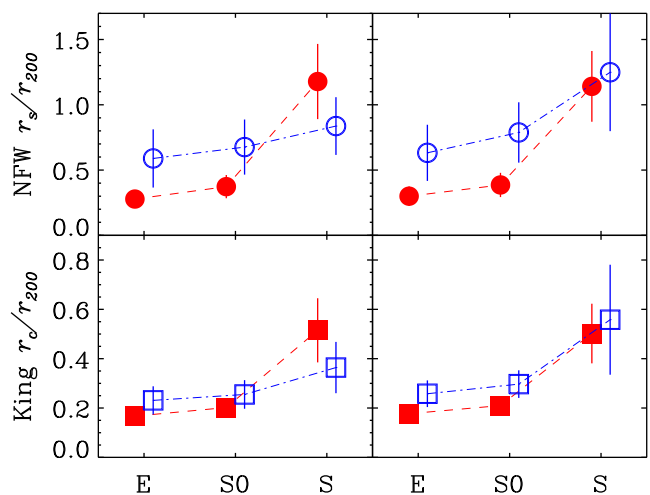}
   \caption{Normalized scale radius as a function of the galaxy
     population for the Reg (red dashed lines and filled symbols) and
     Irr (blue dash-dotted lines and open symbols ) sample. Top panels
     and dots (respectively bottom panels and squares) are for the NFW
     (respectively King) scale-radius. Left (respectively right) panels are for the
     $r_{200,\sigma}$ (respectively $r_{200,N}$) scaling. Error bars are computed 
     according to Eq.\,16 of \cite{BFG90}.}
\label{fig:rhonu}
\end{figure} 

The results of the $\Sigma(R)$ fits are listed in Table~\ref{tab:rnufits},
where we give the values of the average NFW and King scale
radii in units of <$r_{200}$>, as previously explained. 
We find that independent of the model, the scale
radius increases from E to S0 to S (see Table~\ref{tab:rnufits} and
Figure~\ref{fig:rhonu}), as expected from the well-known
morphology-density relation \citep{Dressler80}. This is true both for
Reg and Irr clusters, and for both $r_{200}$ scalings.

We also find that the concentrations, $r_{200}/r_s$ and $r_{200}/r_c$
for the NFW and King models, respectively, of the spatial
distributions of E and S0 galaxies are higher in Reg than in Irr clusters, again
independently of the $r_{200}$ scaling. This is not true for the
concentration of the spatial distribution of S galaxies.  We provide a possible
interpretation for this new result.  Dynamically relaxed cluster-size
halos from cosmological simulations are known to display a higher
concentration (per given mass) than their unrelaxed counterparts
\citep[e.g.,][]{Jing00,Neto+07}. This is probably the consequence of
recent (major) mergers occurring in unrelaxed clusters.  This could
also explain the higher concentration of E and S0 galaxy distributions
in Reg clusters, if these kind of galaxies are good tracers of the
mass distribution.  On the other hand, the insensitivity of the S
spatial distribution to their cluster relaxation state suggests that S
galaxies are recent newcomers in the cluster potential, and that they
might not have settled down in a dynamical equilibrium configuration
yet.

\begin{table}
\centering
\caption{Surface number density profiles of cluster galaxy populations.}
\label{tab:rnufits}
\resizebox{1.\columnwidth}{!}{
\begin{tabular}{llrrr}
\hline 
Stack & Class & NFW <$r_s/r_{200}$> & King <$r_c/r_{200}$> & <$r_{200}$>$^{\rm a}$ \\
\hline
Reg-sigv & E  & $0.278 \pm 0.065$ & $0.167 \pm 0.025$ & $1.749 \pm 0.064$ \\
             & S0 & $0.373 \pm 0.089$ & $0.202 \pm 0.027$ & \\
             & S  & $1.178 \pm 0.288$ & $0.515 \pm 0.130$ & \\
Irr-sigv & E  & $0.589 \pm 0.223$ & $0.231 \pm 0.057$ & $1.976 \pm 0.157$ \\
             & S0 & $0.676 \pm 0.211$ & $0.255 \pm 0.058$ & \\
             & S  & $0.837 \pm 0.221$ & $0.364 \pm 0.104$ & \\
Reg-Num      & E  & $0.300 \pm 0.075$ & $0.177 \pm 0.027$ & $1.629 \pm 0.022$ \\
             & S0 & $0.386 \pm 0.093$ & $0.210 \pm 0.027$ & \\
             & S  & $1.141 \pm 0.271$ & $0.502 \pm 0.121$ & \\
Irr-Num      & E  & $0.632 \pm 0.215$ & $0.258 \pm 0.054$ & $1.652 \pm 0.053$\\
             & S0 & $0.788 \pm 0.231$ & $0.297 \pm 0.056$ & \\
             & S  & $1.249 \pm 0.451$ & $0.558 \pm 0.223$ & \\
\hline
\end{tabular}
} 
\tablefoot{$^{\rm a}$ <$r_{200}$>, in Mpc, is the
weighted average of the individual cluster $r_{200}$ estimates, namely
$r_{200,\sigma}$ for the Reg-sigv and Irr-sigv stacks, and
$r_{200,N}$ for the Reg-Num and Irr-Num stacks.}
\end{table}

\section{The line-of-sight velocity dispersion profiles} 
\label{sec:vdps}

\begin{figure*}
  \centering
\includegraphics[trim=1.8cm 0.4cm 1.6cm 0.cm, clip=true,width=.75\textwidth]{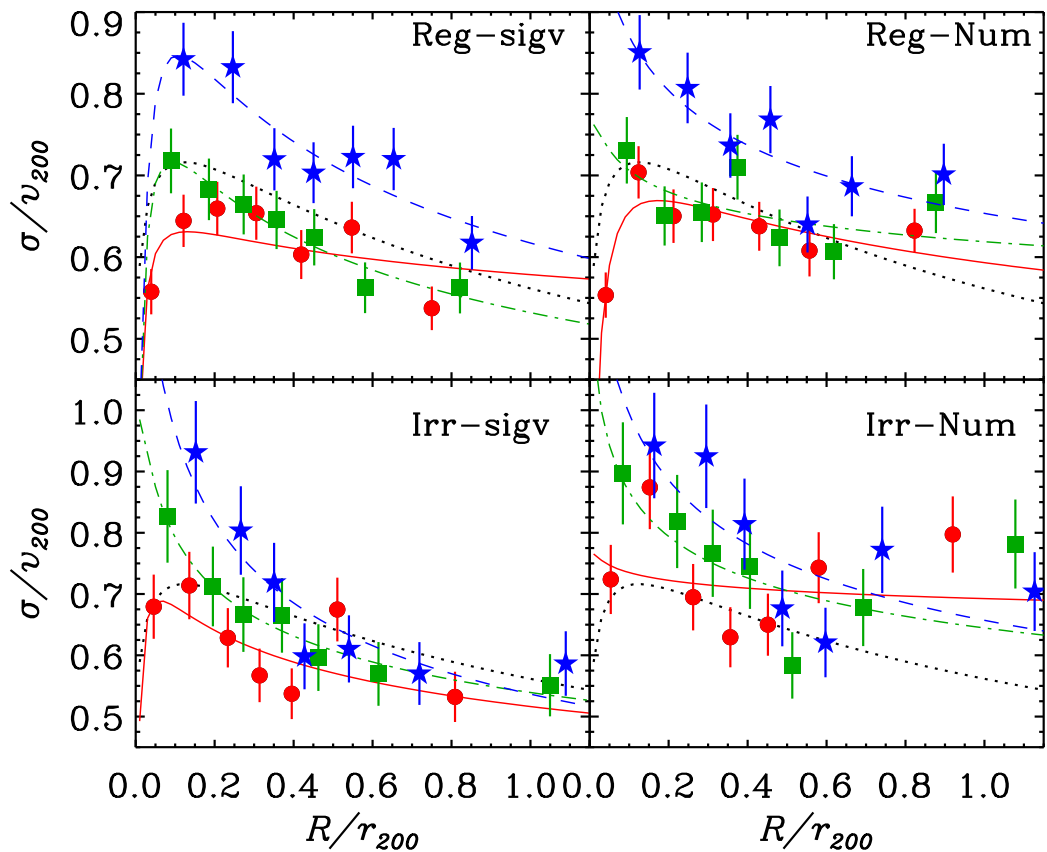} 
     \caption{Line-of-sight velocity dispersion profiles 
       of different cluster galaxy populations in the
       four stack samples (top left: Reg-sigv; top-right: Reg-Num;
       bottom-left: Irr-sigv; bottom-right: Irr-Num). Red dots, green
       squares, and blue stars represent the VDPs of E, S0, and S
       galaxies, respectively. The red solid, green dash-dotted, and
       blue dashed lines represent model fits described by
       Eq.(\ref{eq:vdpmod}) and Table~\ref{tab:vdpmod}. The black
       dotted line represents the theoretical VDP predicted for a
       cluster with a NFW mass distribution of concentration $c \equiv
       r_{200}/r_s=3$ and with a velocity anisotropy profile for its
       galaxies given by Eq.(60) in \citet{ML05b} with
       $r_a=r_s$. Error bars are computed 
     according to Eq.\,16 of \cite{BFG90}}
\label{fig:vdps}
\end{figure*}

We determine the line-of-sight velocity dispersion profiles (VDPs in
the following) of different galaxy populations in our four stack
samples, by computing the biweight \citep{BFG90} velocity dispersion
in concentric radial bins. We need not worry about completeness in
this analysis. In fact, observational selection is unlikely to operate
in velocity space within the (relatively) narrow velocity range
spanned by cluster members.  
So, even if some radial bins lack a higher
fraction of galaxies than others, the lacking galaxies are unlikely to
occupy a preferred location in the global velocity distribution of
cluster members, and therefore the velocity dispersion estimate
remains unaffected by the sample incompleteness.  To confirm this
  expectation we have performed 100 bootstrap simulations, randomly
  extracting from each radial bin a random number of galaxies
  from the observed sample. 
  The relative incompleteness across different radial bins is at most a factor 2. Therefore we choose to simulate the effect of such a random sampling by selecting different fractions of observed galaxies in the different bins, where these fractions are randomly chosen uniformly sampling the range of 50-100\% completeness.
  We confirm that introducing different levels of
  radial completeness in the galaxy distribution does not introduce
  any significant bias in the VDP determination. We therefore use
the full spectroscopic sample for this analysis.

The VDPs are shown in Fig.~\ref{fig:vdps} separately for the E, S0,
and S populations in the four stacks. We used seven bins for each
subsample, with the same number of galaxies per bin. The dotted line
in each panel of the Figure is the predicted VDP ('theoretical VDP' in
the following) for a cluster with a NFW mass distribution of
concentration $c \equiv r_{200}/r_s=3$, typical of massive clusters
\citep[e.g.,][]{GGS16}, and with a velocity anisotropy profile of
cluster galaxies given by Eq.(60) in \citet{ML05b} with anisotropy
radius equal to the NFW mass profile scale radius, $r_a=r_s$
\citep[this anisotropy profile fits cluster-size simulated halos quite
  well,][]{MBM10}.

In Fig.~\ref{fig:vdps} we also show best-fit models to the binned
VDPs of E, S0, and S galaxies. We use the following model,
\begin{equation}
\frac{\sigma_{\rm los}}{v_{200}}=s_0 \, \frac{X}{a_{\sigma}} \, \left( 1+\frac{X}{a_{\sigma}} \right)^{-\eta},
\label{eq:vdpmod}
\end{equation}
where $X=R/r_{200}$, and $s_0$ and $a_{\sigma}$ are in units of
$v_{200}$ and $r_{200}$, respectively.  
We find this model to provide an adequate fit to (almost) all our
VDPs. In Table~\ref{tab:vdpmod}, we provide the $\chi^2$ values of the
fits, together with the best-fit parameters.  Since this model is
a useful parametrization of the cluster VDPs, but has no physical
meaning, we consider it not useful to estimate (and list) the
uncertainties on the fitting parameters.

\begin{table}
\centering
\caption{Best-fit parameters of model VDPs.}
\label{tab:vdpmod}
\resizebox{1.\columnwidth}{!}{
\begin{tabular}{llrrrr}
\hline 
Stack & Class & $s_0$  & $a_{\sigma}$ & $\eta$ & $\chi^2$ \\
      &        & [$v_{200}$] & [$10^{-3} r_{200}$] & & \\
\hline
Reg-sigv & E  &   1.17 &  30.5 &   1.21 &   3.7 \\
             & S0 &   1.20 &  8.4 &   1.16 &   1.8 \\
             & S  &   1.59 &  27.2 &   1.25 &   4.1 \\
Irr-sigv & E  &   1.10 &  10.0 &   1.16 &   7.4 \\
             & S0 &   1.69 &   2.0 &   1.19 &   2.8 \\
             & S  &   2.40 &   4.5 &   1.28 &   4.4 \\
Reg-Num      & E  &   1.02 &  23.8 &   1.14 &   2.9 \\
             & S0 &   1.08 &   0.7 &   1.08 &   3.9 \\
             & S  &   1.46 &   2.8 &   1.14 &   6.4 \\
Irr-Num      & E  &   0.87 &   0.2 &   1.03 &  12.3 \\
             & S0 &   1.75 &   1.5 &   1.16 &   21.0 \\
             & S  &   1.99 &   3.6 &   1.20 &   8.6 \\
\hline
\end{tabular}
} 
\tablefoot{These are the parameters of the model given by
  Eq.(\ref{eq:vdpmod}). The $\chi^2$ values are obtained for the fits
  to seven data-points in each case.}
\end{table}

The comparison of the observed VDPs with the theoretical VDP (dotted
line in Fig.~\ref{fig:vdps}) suggests that our stack clusters should
have an internal structure that is not too dissimilar from the one
assumed to derive the theoretical VDP, at least for a certain class of
galaxies. Perhaps the stack with the VDP that most strongly deviates
from the theoretical VDP is Irr-Num. We postpone a full
dynamical analysis of our data set to Paper II in this series (Mamon
et al., in prep.).

What is more pertinent to this paper is the direct comparison of the
VDPs of the different classes. To perform this comparison, we
re-determine the velocity dispersion of each galaxy population within
fixed radial bins, common to all classes (and all
stacks). Specifically, we use seven radial bins with inner and outer radii
$(i-1) \times 0.15 \, r_{200}$ and $i \times 0.15 \, r_{200}$, for $i=1,7$.
We then evaluate the $\chi^2$ statistic given by
\begin{equation}
  \chi^2 = \sum_{i=1}^7 {\displaystyle
    \frac{\left(\sigma_{x,i}-\sigma_{y,i}\right)^2}{\delta_{x,i}^2+\delta_{y,i}^2}},
  \label{eqvdp}
  \end{equation}
where $\sigma_{x,i}, \sigma_{y,i}$ are the velocity dispersions of
populations $x$ and $y$ in the $i$-th radial bin, and $\delta_{x,i},
\delta_{y,i}$ are their errors. We compare this statistic to a
$\chi^2$ distribution to obtain the probabilities that the two
populations have the same VDPs. These are listed in
Table~\ref{tab:vdpcmp}.

We see that E and S0 galaxies have similar VDPs in all stacks,
although the S0 VDP tends to be steeper than the E VDP. On the other
hand, E and S galaxies have different VDPs in all stacks, and the S VDP is
above the E VDP, especially near the center. In the outer regions, the
S VDP becomes more similar to the E VDP because it is steeper. Our
results are in agreement with the findings of \citet[][see their
  Fig. 9]{ABM98} and consistent with the results of \citet[][see their
  Fig. 2]{BK04}, although the latter did not distinguish E from
S0. However, in addition to these previous studies, here we
distinguish regular from irregular clusters.

\begin{table}
\centering
\caption{Comparison of the cluster galaxy population VDPs.}
\label{tab:vdpcmp}
\resizebox{1.\columnwidth}{!}{
\begin{tabular}{llrrr}
\hline 
Stack & Classes & Prob & \multicolumn{2}{c}{VDP ratio $y$/$x$} \\
      & $x$ vs. $y$ &      & $\leq 0.5 r_{200}$ & $>0.5 r_{200}$ \\
\hline
Reg-sigv & E  vs. S0 & 0.184 & $1.08 \pm 0.08$ & $ 1.04 \pm 0.09$ \\
             &\bf E  vs. S  & $\bf <0.001$ & $\bf 1.27 \pm 0.11$ & $\bf 1.22 \pm 0.06$ \\
             &\bf S0 vs. S  & $\bf <0.001$ & $\bf 1.17 \pm 0.04$ & $\bf 1.17 \pm 0.04$ \\
Irr-sigv & E  vs. S0 & 0.125 & $1.15 \pm 0.07$ & $ 0.98 \pm 0.10$ \\
             &\bf E  vs. S  &\bf 0.010 & $\bf 1.35 \pm 0.07$ & $\bf  1.03 \pm 0.04$ \\
             & S0 vs. S  & 0.194 & $1.08 \pm 0.10$ & $ 1.06 \pm 0.11$ \\
Reg-Num      & E  vs. S0 & 0.287 & $1.04 \pm 0.08$ & $ 0.96 \pm 0.04$ \\
             &\bf E  vs. S  & $\bf<0.001$ & $\bf1.24 \pm 0.06$ & $\bf 1.12 \pm 0.02$ \\
             &\bf S0 vs. S  &\bf 0.008 & $\bf1.11 \pm 0.00$ & $\bf 1.09 \pm 0.04$ \\
Irr-Num  & E  vs. S0 & 0.084 & $1.15 \pm 0.11$ & $ 0.94 \pm 0.09$ \\
             &\bf E  vs. S  &\bf 0.017 & $\bf1.26 \pm 0.05$ & $\bf 0.94 \pm 0.02$ \\
             & S0 vs. S  & 0.150 & $1.12 \pm 0.07$ & $ 0.91 \pm 0.15$ \\
\hline
\end{tabular}
} 
\tablefoot{"Prob" indicates the probability that the VDPs of the x and y
  populations are the same. We consider probabilities $<0.05$ as
  indicating a significant difference in the VDPs. 
  These cases are highlighted in boldface in the table.
  The VDP ratios are computed in two radial bins.}
\end{table}

An entirely new result of our analysis is how the S0 galaxies' VDP compares to
the VDPs of the other two classes in Reg and Irr clusters. In Reg
clusters, the S0 VDP is significantly different from the S VDP, being
systematically below at all radii.  In Irr clusters, the S0 VDP
remains below the S VDP at all radii, but the difference is no longer
significant. It therefore appears that while S0 galaxies in Reg
clusters have similar velocity distributions to E galaxies, in Irr
clusters, the velocity distribution of S0 galaxies is intermediate between that
of S and E.

The different behavior of the S0 VDP with respect to the other classes'
VDPs in Reg and Irr clusters suggests that S0 galaxies in Irr clusters are
a younger population than S0 galaxies in Reg clusters. In Irr clusters,
relatively to Reg clusters, a larger fraction of S0 galaxies is likely to have 
relatively recently evolved from infalling S galaxies.

\section{Discussion}
\label{sec:disc}
In Sect.~\ref{sec:stacks} we have shown that the PPSDs
(for all the considered stacks and scalings) are statistically
different for the three classes of galaxies considered in this study:
E, S0, and S. This result is at variance with that of
\citet{Biviano+02}, based on the European Southern Observatory Nearby Abell Cluster Survey
\citep[ENACS;][]{Katgert+96,Katgert+98}, where no statistical evidence was found
for different PPSDs of E and S0. We argue that this might
be due to \citet{Biviano+02} using the mixed morphological and
spectroscopical classification of galaxies by \citet{TK06a}, which
differs from the pure morphological classification used in this
work. It is plausible that a substantial fraction of
spectroscopically-classified S0 galaxies are in fact morphological E
galaxies \citep{TK06a}, and this contamination would reduce the
detectability of any intrinsic difference in the distributions of the
two populations.

In Sects.~\ref{sec:profs} and \ref{sec:vdps} we have dissected the PPSDs 
into their spatial and velocity components. The spatial
distributions were parametrized by the scale- or core-radius of the
best-fitting NFW or King models to the surface number density profiles
of cluster galaxies. We find that in general the NFW profile is
preferred over the cored-King profile, but this is less and less
evident for late-type galaxies and Irr clusters.  The spatial
distributions become increasingly more concentrated towards earlier
morphological classes, and this reflects the well-known
morphology-radius relation \citep[][and references
  therein]{SS90b,WGJ93,Fasano+15}.  The velocity distributions were
described in terms of the velocity dispersion profiles, that we
modeled with the analytical profile of Eq.~\ref{eqvdp}. The velocity
dispersion profile has a higher normalization and is steeper for later
morphological classes, confirming previous results \citep{ABM98,BK04}.

These morphological segregations in the spatial and velocity
distributions have been generally interpreted as E being the oldest
cluster population, followed by S0, and then S
\citep[e.g.,][and references therein]{Dressler+97,BK03,Biviano11}.
The high-$z$ cluster
environment is believed to have favored the morphological evolution of
field galaxies into E, presumably via mergers, before the cluster
velocity dispersion grew too high for mergers to be effective.  Later
evolution concerns the transformation of field S into cluster S0, and
there is direct evidence that the fraction of cluster S decreases with
time since $z \sim 1$, at the same time as the fraction of cluster S0
increases \citep{Postman+05,Smith+05,Desai+07}.

We have found that the distributions of E, S0, and S galaxies depend on the
kind of clusters they belong to.  In particular, both E and S0 have
more concentrated spatial distributions (i.e., greater $r_{200}/r_s$ or
$r_{200}/r_c$) in Reg than in Irr clusters. This is not the case for
S.

Dynamically relaxed cluster-size halos from cosmological simulations
have higher mass density concentrations than unrelaxed halos of the
same mass \citep[e.g.,][]{Jing00,Neto+07}. The fact that we see the
same trend observationally in the distribution of E and S0, but not S,
suggests that E and S0 galaxies are good tracers of the cluster mass
distribution. S galaxies are not, probably because they are a recently
accreted population in both Reg and Irr clusters, while (most)
E and S0 galaxies have resided in the cluster environment for a sufficiently
long time to settle down into a spatial distribution similar to that of the
whole cluster mass.

Another important difference between Reg and Irr clusters is evidenced
in the VDP of S0. This appears very similar to the VDP of E galaxies in Reg
clusters, and intermediate between those of E and S in Irr clusters.
S0 galaxies share the velocity distribution of E in the more evolved Reg clusters,
presumably because they have had the time to settle down in the
cluster potential, unlike in the less evolved Irr clusters.
The average time since morphological evolution of S0 galaxies from infalling S galaxies must
be shorter in Irr than in Reg clusters. It is possible that this
difference in age could be visible in the S0 spectra, but such an analysis is
beyond the scope of this work.

The different spatial and velocity distributions of S0 and S galaxies suggest
that it is the cluster environment that operates the transformation
process between these two morphological types. In fact, if most S0 galaxies evolve from S galaxies in groups before being accreted onto clusters
\citep[the so-called 'pre-processing';][]{ZM98art,Balogh+99,Mahajan13}, S and S0
should share similar spatial and velocity distributions. 
S0 galaxies are therefore expected to be the result of a transformation process
of S galaxies operated by the cluster environment. 
The different PPSD of S0 and S galaxies might result from
the fact that the morphological evolutionary processes are more effective
in the central cluster regions \citep[e.g.,][]{Moran+07}. 
In particular, ram-pressure stripping \citep{GG72}, 
and/or starbursts induced by tidal compression by the cluster gravitational field,
especially while the galaxies are accreted in groups \citep[the so-called 'post-processing';][]{Bekki99,Oemler+09,VR13,Stroe+15,Jaffe+16}, 
likely play a relevant role. This compression could drive the rapid consumption of gas funneled
into the central galaxy region and the build-up of a central bulge.

Previous works have indicated that $\geq 2/3$ of cluster S0 form after
$z \sim 0.4$, that is, in the last 3-4 Gyr
\citep{Postman+05,Smith+05,Desai+07}. This sets an upper limit to the
timescale for morphological evolution. We also observe that S0 and S
have different PPSDs, so also this differentiation must
occur on a timescale $\lesssim 3$ Gyr. This upper limit is consistent
with the dynamical timescale, $t_{dyn} \sim (G <\rho>)^{-1/2}$
\citep{Sarazin86}, that is $\approx 2$ Gyr in the virialized cluster
region.\footnote{More precisely, the orbital time in a cluster is $\sqrt{8}\pi~(\sqrt{\Delta}~H)^{-1}$ ($\approx8.9~(\sqrt{\Delta}~H)^{-1}$) for circular orbits
and $\approx7~(\sqrt{\Delta}~H)^{-1}$ for radial orbits (for NFW profiles, 
where $\Delta$ is the mean cluster overdensity relative to critical at the apocenter of the orbit). 
This leads to free-fall times from the virial radius of 2.1~Gyr at $z=0.25$ (3~Gyr ago) and 
2.6~Gyr now (for reasonably elongated orbits).}

'Delayed quenching' models suggest that S can live in the
cluster environment for several Gyr before being quenched
\citep{Wetzel+13,Fossati+17}. The inferred timescales for
morphological and PPS evolution from our and previous analyzes set an
upper limit to the delayed quenching models timescale of $2-3$ Gyr,
and cannot exclude more rapidly quenching models \citep{OH16}.

\section{Conclusions}
\label{sec:concl}
We use the WINGS data set to investigate the spatial and velocity
distribution of cluster galaxies of three morphological classes, E,
S0, and S. We do so by building stack clusters, using two different
stacking normalization parameters, one based on velocity dispersion,
the other on cluster richness. Our results do not depend on the
stacking parameter.  We build two stacks, one for regular (Reg)
clusters and another for (Irr) clusters with significant evidence for
substructures according to the DS test.

We determine the surface number density and velocity dispersion
profiles $\Sigma(R)$ and $\sigma(R)$, of cluster E, S0, and S galaxies to an
unprecedented accuracy. We fit $\Sigma(R)$ with both NFW and King
models, and obtain marginally better fit results for the NFW model,
independently of the morphological class and in both Reg and Irr
clusters.  We propose a new phenomenological model that provides
acceptable fits to the $\sigma(R)$ of all our galaxy samples.

These are our main results.
\begin{itemize}
\item The projected phase space distributions of E, S0, and S galaxies are all different, both in Reg and in Irr clusters, at variance with
  a previous result \citep{Biviano+02}.
\item The concentration, $r_{200}/r_s$ or $r_{200}/r_c$,
  increases from E to S0 to S galaxies, both in Reg and Irr clusters,
  reflecting the well-known morphology-radius relation \citep{SS90b,WGJ93,Fasano+15}.
\item The concentration is higher in Reg
  than in Irr clusters for both E and S0 galaxies, but not for spirals.
\item The velocity dispersion profiles have increasingly higher normalization and
  steeper slope for increasingly later-type galaxies (from E to S0 to S types),
  in agreement with the results of \citet{ABM98,BK04}.
\item In Reg clusters, E and S0 galaxies have similar $\sigma(R)$, different from
  that of spirals; in Irr clusters, the S0 velocity dispersion profile is intermediate between
  those of the other two classes.
\end{itemize}

Our results indicate that the cluster environment is driving the
transformation from infalling spirals to a dynamically more relaxed S0
population. In combination with previous results on the evolving
fraction of cluster S0 galaxies with time \citep{Postman+05,Smith+05,Desai+07},
we argue that the timescale for this transformation is $\leq 3$ Gyr.
This transformation appears to be mostly completed in nearby Reg
clusters, but still ongoing in nearby Irr clusters. We therefore
expect S0s in Irr clusters to have lower ages on average than those in Reg
clusters. Furthermore, we expect the PPSD of S0s in $z
\gtrsim 0.4$ clusters, at a time where most S0s should be still
evolving from infalling spirals, to be more similar to that of spirals than to
that of ellipticals, unlike what we see in nearby clusters. Future investigations
will be able to confirm or reject our expectations.

In future papers of this series we will use some of the results of
this paper to perform a full dynamical analysis of the WINGS data-set,
and to determine, in particular, the mass and velocity anisotropy
profile of Reg WINGS clusters (Paper II: Mamon et al. in prep.) and their
pseudo phase-space density profile (Paper III: Biviano et al. in prep.).

\begin{acknowledgements} 
We wish to thank the anonymous referee for her/his useful comments that have helped to improve this manuscript.
The work of AC is supported by the STARFORM Sinergia Project funded by
the Swiss National Science Foundation. This research has made use of
the X-Rays Clusters Database (BAX) which is operated by the
Laboratoire d'Astrophysique de Tarbes-Toulouse (LATT), under contract
with the Centre National d'Etudes Spatiales (CNES). This research has
made use of the NASA/IPAC Extragalactic Database (NED) which is
operated by the Jet Propulsion Laboratory, California Institute of
Technology, under contract with the National Aeronautics and Space
Administration. AB acknowledges the hospitality of the Institut
d'Astrophysique de Paris. The work of AB has been financially
supported in part by the PRIN INAF 2014: “Glittering kaleidoscopes in
the sky: the multifaceted nature and role of Galaxy Clusters”, P.I.:
Mario Nonino, by NSF grant AST-1517863. 

\end{acknowledgements}

\bibliographystyle{aa} 
\bibliography{paper1} 

\end{document}